\documentclass[5p]{elsarticle}
\usepackage{amsmath,amssymb}
\usepackage{color}
\usepackage{tabularx}
\usepackage{lineno}\modulolinenumbers[5]

\usepackage[euler]{textgreek}
\usepackage{siunitx}
\DeclareSIUnit{\wtpercent}{wt.\%} 
\DeclareSIUnit{\atpercent}{at.\%} 
\usepackage{miller}

\journal{Scripta Mater.}

\begin{document}

\begin{frontmatter}

\title{Crystallographic ordering of Al and Sn in \textalpha{}-Ti}

\author[IC]{Felicity F. Worsnop}
\author[IC]{Susannah L. M. Lea}
\author[Argonne]{Jan Ilavsky}
\author[RR]{David Rugg}
\author[IC]{David Dye}

\address[IC]{Department of Materials, Royal School of Mines, Imperial College London, Prince Consort Road, London, SW7 2BP, UK}
\address[Argonne]{Advanced Photon Source, Argonne National Laboratory, 9700 S Cass Avenue, Argonne, IL, USA}
\address[RR]{Rolls-Royce plc., Elton Road, Derby, DE24 8BJ, UK}

\begin{abstract}
Increasing attention is being paid to $\alpha_2$ Ti$_3$(Al,Sn) precipitation from the $\alpha$ phase of titanium alloys owing to its effect on slip band formation, localisation and the implications for fatigue performance in jet engine titanium. However, the early stages of $\alpha_2$ precipitation have historically been difficult to observe in TEM, neutron diffraction or atom probe analysis. Here, small angle X-ray scattering is used to reexamine the phase boundary in binary Ti-Al and Ti-Sn alloys with around 500 ppmw O. It is found that the phase boundaries in the literature are approximately correct, at 6.2 wt.\% Al and 16.9 wt.\% Sn, and that this favours the use of Al as a solid solution strengthener over Sn for ambient temperature applications. However, once O content and phase partitioning in \textalpha{}+\textbeta{} alloys are taken into account, this implies that Al$_{\mathrm{eq}}$ limits for future alloy design of critical rotating parts should be lowered substantially.
\end{abstract}


\end{frontmatter}

Many \textalpha{}+\textbeta{} titanium alloys use Al and Sn to stabilise and strengthen the hcp \textalpha{} phase, producing materials that have excellent property combinations for aerospace and industrial applications. Sn in particular is included for retention of strength at elevated temperatures. However, these solutes are known to undergo crystallographic ordering after exposure to temperatures experienced during processing, forming the \textalpha{}$_2$ Ti$_3$(Al,Sn) phase \cite{Murray1990}. For instance, Ti--834 (Ti--5.8Al--4Sn--3.5Zr--0.7Nb--0.5Mo--0.35Si--0.06C) produces a fine dispersion of \textalpha{}$_2$ precipitates after \SI{24}{\hour} at \SI{700}{\celsius} \cite{LandW}. This ordered phase promotes slip localisation that increases the risk of dwell fatigue crack nucleation in critical rotating parts of jet engines. Under stress corrosion cracking conditions, crystallographic ordering may arise around the crack front as a result of corrosion reactions \cite{Shi2021}. Improved understanding of propensity to ordering for these solutes could aid alloy and processing design to mitigate potential failure mechanisms.

Here, binary Ti--Al and Ti--Sn alloys were investigated to compare their tendency for crystallographic ordering during ageing at \SI{550}{\celsius}. The \textalpha{} and \textalpha{}$_2$ microstructures were characterised with SEM, TEM and small angle X-ray scattering (SAXS) alongside tensile testing to reconsider the property trade-offs in each system.


Binary compositions were chosen at and either side of the \textalpha{}/\textalpha{}+\textalpha{}$_2$ boundary, based on the chosen ageing temperature and currently accepted phase diagrams \cite{Murray1990,Schuster2006}. Ingots were arc melted under argon from pure elements, then processed to achieve equiaxed \textalpha{} microstructures using conditions appropriate for each composition, Table~\ref{table:composition-processing}. Bars were ice water quenched from the recrystallisation temperature, and half of each bar was then aged at \SI{550}{\celsius} under argon for \SI{80}{\day}. Compositions were measured using ICP-OES and LECO at TIMET UK, Witton, UK.

\begin{table*}[t!]
	\centering
	\caption{Compositions, processing details and grain sizes for the Ti--Al and Ti--Sn alloys. Rolling and recrystallisation temperatures were chosen according to each alloy's \textbeta{} transus. The \textbeta{} rolling temperature was \SI{1150}{\celsius} for Ti--Al alloys and \SI{1050}{\celsius} for Ti--Sn alloys. All materials were ice water quenched (IWQ) after solutionising at the recrystallisation temperature. Grain sizes were measured using the mean linear intercept method (per ASTM E112 \cite{ASTME112}).}
	\begin{tabular}{l | c c c c | c | c c c | c} 
	\hline
	Alloy (nominal & \multicolumn{4}{c|}{Measured composition / wt.\%} & at.\% Al or Sn& $T_{\mathrm{roll}}$ / \si{\celsius} & $T_{\mathrm{RX}}$ / \si{\celsius} & $t_{\mathrm{RX}}$ / h & Grain size / \si{\micro\metre}\\
	wt.\%) & Al & Sn & O & N & & & & \\
	\hline
	Ti--3.5Al & 3.55 & $<$0.01 & 0.04 & 0.01 & 6.12 & 980 & 900 & 1 & 50\\
	Ti--5.3Al & 5.28 & $<$0.01 & 0.03 & 0.01 & 8.99 & 980 & 900 & 2 & 49\\
	Ti--7Al & 7.00 & $<$0.01 & 0.08 & 0.01 & 11.76 & 980 & 900 & 3 & 57\\
	Ti--7Sn & $<$0.01 & 7.10 & 0.07 & 0.01 & 2.98 & 800 & 780 & 15 & 19\\
	Ti--12Sn & $<$0.01 & 12.02 & 0.06 & 0.01 & 5.21 & 800 & 780 & 12 & 20\\
	Ti--17Sn & $<$0.01 & 17.00 & 0.04 & $<$0.01 & 7.62 & 800 & 780 & 4 & 13\\
	\hline
	\end{tabular}
	\label{table:composition-processing}
\end{table*}

Microstructures were characterised using backscatter SEM (Zeiss Sigma 300, 8--\SI{12}{\kilo\volt}) and conventional TEM electron diffraction and dark field imaging (JEOL 2100F, \SI{200}{\kilo\electronvolt} accelerating voltage, with double-tilt specimen holder). Specimens were electropolished with a solution of 3\% perchloric acid, 40\% butan-1-ol and 57\% methanol, using an applied voltage of 20.0--\SI{21.5}{\volt} at \SI{-45}{\celsius} to \SI{-30}{\celsius}. Tensile tests were performed with an Instron 5967 load frame and extensometer at a nominal strain rate of $10^{-3}$~s$^{-1}$, using a flat dogbone geometry (\SI{19}{\milli\metre}$\times$\SI{1.5}{\milli\metre}$\times$\SI{1.5}{\milli\metre} gauge).

SAXS measurements were taken on the USAXS beamline \cite{Ilavsky2018} at the Advanced Photon Source at Argonne National Laboratory, using a \SI{21}{\kilo\electronvolt} beam and a 800$\times$\SI{800}{\micro\metre} beam area. Samples were prepared by grinding to \SI{300}{\micro\metre} thickness with SiC paper from 500 to 4000 grits, followed by neutralised colloidal silica solution. After cleaning with detergent and isopropanol, samples were held in amorphous tape during measurement. Data were reduced and analysed using the Nika and Irena packages \cite{Ilavsky2012,Ilavsky2009}, informed by our previous analysis of \textalpha{}$_2$ composition \cite{Dear2021}.

\begin{figure*}[h!]
   \centering
   \includegraphics[width=1.0\textwidth]{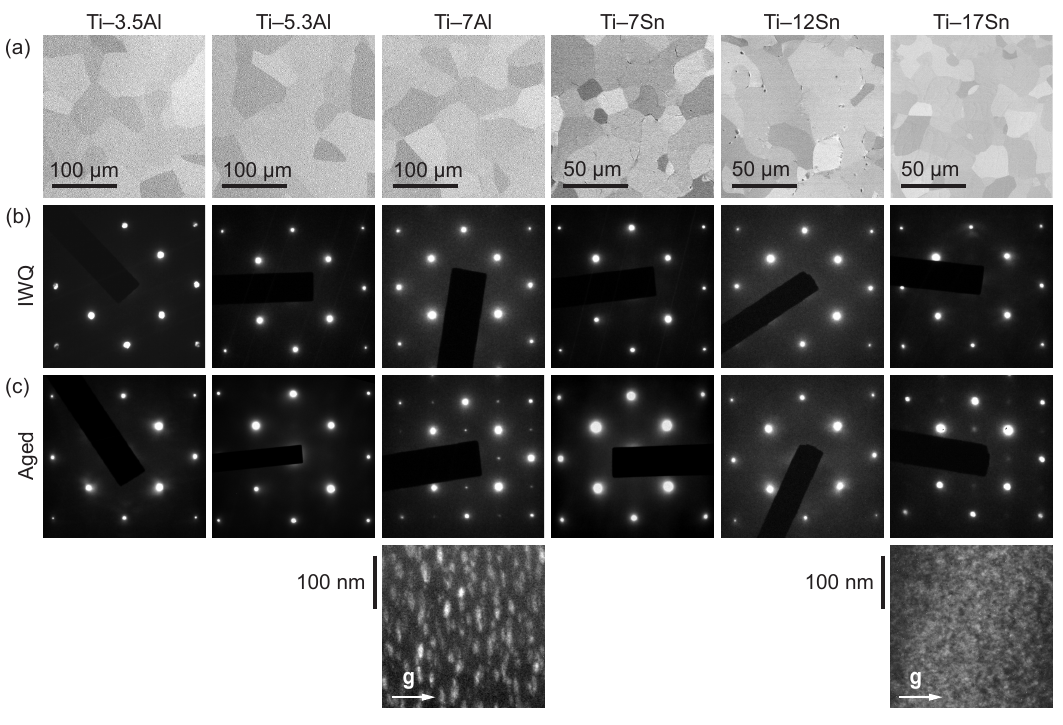}
   \caption{Microstructures of the Ti--Al and Ti--Sn binary alloys. (a) Backscatter SEM showing equiaxed \textalpha{} grains. The Ti--Sn alloys were found to be susceptible to hydride formation during electropolishing, visible at some grain boundaries in SEM. (b) Selected area diffraction patterns taken along \textbf{B}~=~\hkl<0 1 -1 1>. (c) Dark field images were recorded using a \hkl<2 -1 -1 0>$_{\alpha{}2}$ two-beam condition with \textbf{B}~=~\hkl<0 1 -1 1>.}
   \label{fig:microstructures}
\end{figure*}

Equiaxed, single-phase \textalpha{} microstructures were produced for each composition, Fig.~\ref{fig:microstructures}, with grain sizes of \SI{50}{\micro\metre} and \SI{20}{\micro\metre} in the Ti--Al and Ti--Sn alloys respectively, Table~\ref{table:composition-processing}.

The IWQ and aged conditions were compared using selected area diffraction and dark field imaging in TEM for a qualitative indication of whether \textalpha{}$_2$ formation had occurred. No superlattice reflections were observed for any of the alloys in the IWQ state, suggesting suppresion of long-range ordering during the quench from \SI{900}{\celsius} (Ti--Al) or \SI{780}{\celsius} (Ti--Sn).

In the Ti--Al series, only Ti--7Al developed superlattice reflections, and dark field images showed well formed precipitates. No contrast could be observed in attempts to take dark field images for aged Ti--3.5Al or Ti--5.3Al. In the Ti--Sn series, ageing produced faint superlattice reflections for the highest solute content only, Ti--17Sn, which contained a much finer precipitate dispersion than Ti--7Al.

Precipitate aspect ratios were obtained for use in SAXS data fitting. For Ti--7Al, multiple separate particles were measured to give an average aspect ratio of 2.4, while a value of 1.0 was assigned to the Ti--17Sn alloy's precipitates based on their equiaxed appearance.

\begin{figure}[h!]
   \centering
   \includegraphics[width=\columnwidth]{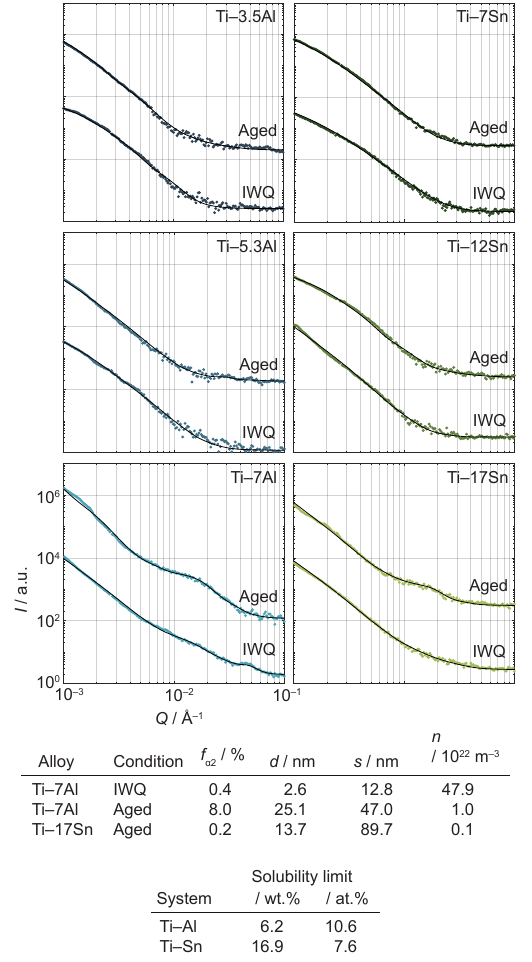}
   \caption{SAXS curves and models for the Ti--Al and Ti--Sn binary alloys, IWQ and aged. The presence of \textalpha{}$_2$ precipitates was identified in Ti--7Al and Ti--17Sn (wt.\%). Below are shown the SAXS-derived volume fraction ($f_{\alpha{}2}$), size ($d$), spacing ($s$) and number density ($n$) for each precipitate population, and the solubility limit (i.e. position of the \textalpha{}/\textalpha{}+\textalpha{}$_2$ boundary on the Ti--X binary equilibrium phase diagram) for each system.}
   \label{fig:saxs}
\end{figure}

Further characterisation of precipitate populations was conducted using SAXS, Fig.~\ref{fig:saxs}. Data were fitted assuming a microstructure of \textalpha{} grain boundaries (modelled as thin discs) and \textalpha{}$_2$ precipitates (modelled as spheroids with aspect ratios informed by TEM). Electron density contrast between \textalpha{} and \textalpha{}$_2$ was calculated based on a stoichiometric Ti$_3$(Al,Sn) phase, giving contrast values of \SI{2.2e20}{\centi\metre\tothe{-4}} for the Ti--Al alloys and \SI{53.6e20}{\centi\metre\tothe{-4}} for the Ti--Sn alloys.

In Ti--3.5Al and Ti--5.3Al, both the IWQ and aged conditions gave no indication of \textalpha{}$_2$ Ti$_3$Al formation, with a model describing the grain boundaries sufficient for fitting in each case. In the Ti--7Al material, the small peak at approximately \SI{4e-2}{\per\angstrom} was modelled as a population of equiaxed \textalpha{}$_2$ Ti$_3$Al nuclei, of mean diameter \SI{3.2}{\nano\metre} and with 0.4\% volume fraction. This may indicate that the quenching in this alloy was insufficient to suppress this early stage of the phase separation process. Upon ageing, the peak shifted to lower $Q$ and a volume fraction of 8.0\% of particles with mean diameter \SI{25}{\nano\metre} was obtained through fitting spheroids of aspect ratio 2.4.

Similarly, in Ti--7Sn and Ti--12Sn, both IWQ and aged conditions showed no indication of precipitation.  Ti--17Sn was also found to have no second phase in the IWQ condition, but a clear peak had developed for the aged specimen, assumed to be the \textalpha{}$_2$ Ti$_3$Sn phase identified in TEM. This was fitted to give a volume fraction of 0.2\%, and a particle diameter of \SI{13.7}{\nano\metre}.

\begin{figure}[t!]
   \centering
   \includegraphics[width=\columnwidth]{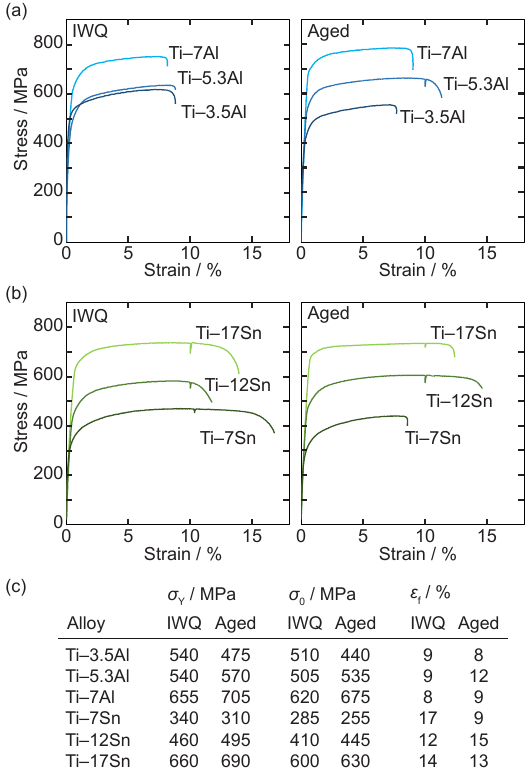}
   \caption{Tensile testing of (a) the Ti--Al binary alloys and (b) the Ti--Sn binary alloys. (c) Resulting yield stresses $\sigma{}_{\mathrm{Y}}$, calculated Hall--Petch intrinsic strengths $\sigma{}_{\mathrm{0}}$ (to discard the effect of grain size), strains at failure $\epsilon{}_{\mathrm{f}}$, and the measured grain sizes used for Hall--Petch analysis.}
   \label{fig:tensile}
\end{figure}

The yield stresses, $\sigma_{\mathrm{Y}}$, and strains at failure, $\epsilon_{\mathrm{f}}$, were obtained from tensile tests of each material, Fig.~\ref{fig:tensile}(a--b). The Hall--Petch relation was used to distinguish the contributions of solute content, ageing state and grain size: \[\sigma_{\mathrm{Y}} = \sigma_{\mathrm{0}} + \frac{k_{\mathrm{HP}}}{\sqrt{D}},\]
where $\sigma_{\mathrm{0}}$ is a material constant, $k_{\mathrm{HP}}$ is the material's Hall--Petch coefficient (an empirically derived constant), and $D$ is the average grain diameter. Here, $\sigma_{\mathrm{0}} = \sigma_{\mathrm{Ti}} + \Sigma{}c_{i}\sigma_{\mathrm{i}}$, where $\sigma_{\mathrm{Ti}}$ is the strength of pure Ti, $c_{i}$ is the concentration of solute $i$ and $\sigma_{\mathrm{i}}$ is the strengthening contribution per wt.\% solute $i$. This applies for the IWQ condition; there is an additional contribution from precipitation strengthening where \textalpha{}$_2$ is present. We took $k_{\mathrm{HP}}$ = \SI{230}{\mega\pascal\micro\metre}$^{-1/2}$ for all materials, the reported value for both CP Ti \cite{LandW} and for equiaxed Ti--64 \cite{Chong2019}.

First considering ageing effects, $\sigma_{0}$ was calculated for each material and compared in order to disregard grain size effects, Fig.~\ref{fig:tensile}(c). The Ti--Al alloys then showed the expected trends with solute content and ageing. In the IWQ state, Ti--3.5Al and Ti--5.3Al had similar $\sigma_{0}$, while Ti--7Al showed a notably higher $\sigma_{0}$. This is attributed to solid solution strengthening as well as the trace \textalpha{}$_2$ detected in SAXS for quenched Ti--7Al. After ageing, Ti--5.3Al and Ti--7Al saw an increase in $\sigma_{\mathrm{0}}$ of \SI{30}{\mega\pascal} and \SI{50}{\mega\pascal} respectively. This is attributed to \textalpha{}$_2$ growth in Ti--7Al and to possible short-range order in Ti--5.3Al. For Ti--3.5Al, slight softening was observed, which we attribute to annealing of dislocation density remaining after processing and the absence of ordered domains.

For the Ti--Sn alloys, there was a clear trend of increasing yield stress with increasing alloying content, in both the IWQ and aged conditions. The Ti--12Sn and Ti--17Sn alloys both gained around \SI{30}{\mega\pascal} in yield stress upon ageing. The Ti--7Sn alloy showed slight softening upon ageing; as for the Ti--3.5Al alloy, we attribute this to a lack of \textalpha{}$_2$ precipitation and annealing out of defect density. For Ti--17Sn, loss of work hardening ability is seen upon ageing (evident as flattening of its stress--strain curve compared to the IWQ condition), consistent with the slip band formation mechanism that is associated with \textalpha{}$_2$ \cite{Brandes2010}. For Ti--7Al, the effect is subtler and is mostly seen as a shortening of the yield region of the stress--strain curve.


Based on the above, values were calculated and compared for $\sigma{}_{\mathrm{Al}}$ and $\sigma{}_{\mathrm{Sn}}$ using the IWQ data, taking $\sigma_{\mathrm{Ti}}$ = \SI{140}{\mega\pascal} \cite{ASMHandbook}. For Ti--Al, $\sigma{}_{\mathrm{Al}}$ = \SI{61}{\mega\pascal}/\si{\wtpercent}~Al, equivalent to \SI{38}{\mega\pascal}/\si{\atpercent}~Al. For Ti--Sn, $\sigma{}_{\mathrm{Sn}}$ = \SI{30}{\mega\pascal}/\si{\wtpercent}~Sn, or \SI{64}{\mega\pascal}/\si{\atpercent}~Sn.

For the compositions that underwent \textalpha{}$_2$ formation, age hardening effects may also be compared. A $\sigma_{\mathrm{0}}$ increase of \SI{55}{\mega\pascal} for Ti--7Al was found, in which an 8\% volume fraction of \textalpha{}$_2$ was produced. In the Ti--17Sn material, ageing produced a low volume fraction (0.2\%) of \textalpha{}$_2$, with a resulting small increase of $\sigma_{\mathrm{0}}$ by \SI{30}{\mega\pascal}.

Orowan strengthening is approximately given by $\Delta\tau_\mathrm{bowing} = Gb \sqrt{(3f/2\pi)}$/r, where $G$ is the shear modulus, $b$ the Burgers vector and $r$ the precipitate radius \cite{Martin}. From the SAXS analysis, the ratio of expected precipitate hardening contributions given by $\sqrt{f}/r$ was 3.5, whereas the tensile test measurements showed that the Ti--7Al alloy had a 1.8$\times$ greater increase in yield strength on ageing than Ti--12Sn. Given the uncertainties in the analyses, and the removal of solute from the matrix (non-diluteness), this order-of-magnitude agreement is felt to provide confidence to the SAXS analysis.

For assessment of the Ti--Al and Ti--Sn binary equilibrium phase diagrams, the lever rule was used to calculate the position of the \textalpha{}/\textalpha{}+\textalpha{}$_2$ boundary at \SI{550}{\celsius} in each system. For a two-phase system, $f_{1}(c-c_{1}) = f_{2}(c-c_{2})$, where $c$ is the alloy composition, $c_1$ and $c_2$ are the compositions of phases 1 and 2, and $f_1$ and $f_2$ are the volume fractions of each phase. The composition at the \textalpha{} single-phase boundary is then $c_{\alpha} = (c-f_{\alpha{}2}c_{\alpha{}2})/f_{\alpha}$. For Ti--7Al and Ti--17Sn, Ti$_3$X stoichiometry was assumed for \textalpha{}$_2$. The solubility limit at \SI{550}{\celsius} was found to be \SI{10.6}{\atpercent} or \SI{6.2}{\wtpercent}~Al in the Ti--Al system, and \SI{7.6}{\atpercent} or \SI{16.9}{\wtpercent}~Sn in the Ti--Sn system, in close agreement with previous studies \cite{Murray1990,Namboodhiri1983}.

This measurement of solubility limits may be used to reassess the widely used metric of aluminium equivalent, which describes an alloy's overall likelihood of straying into the two-phase \textalpha{}+\textalpha{}$_2$ field. The principle, initially put forward by Rosenberg et al. \cite{Rosenberg1968}, calculates a value $\mathrm{Al}_{\mathrm{eq}} = \sum_{i}x_{i}c_{i}$, where $c_i$ are wt.\% compositions of each alloying element, and $x_i$ are empirically derived constants, calculated as the ratio of the solubility limit of element $i$ to that of Al. Current literature gives this as $\mathrm{Al}_{\mathrm{eq}}$ = [Al] + 0.33[Sn] + 0.17[Zr] + 10[O]. Based on the present results, however, the prefactor for Sn might be modified from 0.33 to \SI{0.36}{\per\wtpercent}, i.e. a 10\% increase in its value.


Multiple properties must be considered together to evaluate alloy suitability for a given application, such as strength, density, fatigue resistance, corrosion and oxidation resistance, and the retention of properties at elevated temperature. Binary alloys at the solubility limit in each system, Ti--6.2Al and Ti--16.9Sn (wt.\%) were considered for comparison. Their densities were estimated as \SI{4.30}{\gram\per\centi\metre\cubed} and \SI{5.01}{\gram\per\centi\metre\cubed} respectively, assuming the same lattice parameters as for pure Ti. Interpolated $\sigma_{\mathrm{0}}$ values were found to be \SI{571}{\mega\pascal} and \SI{590}{\mega\pascal} respectively. For a \SI{20}{\micro\metre} grain size, this predicts specific strengths of \SI{145}{\mega\pascal\per\gram\centi\metre\cubed} for Ti--6.2Al and \SI{128}{\mega\pascal\per\gram\centi\metre\cubed} for Ti--16.9Sn, a 13\% benefit for Al over Sn.

For rotating parts of jet engines, fatigue and high-temperature properties are also important. Crystallographic ordering is linked to increased dwell sensitivity, due to micromechanical stress localisation resulting from slip band formation \cite{Brandes2010,Neeraj2001}. Meanwhile, higher \textalpha{}-stabiliser content (particularly Sn) is partly motivated by \textbeta{} fraction reduction for high-temperature alloys, since the higher diffusion rates in \textbeta{} make it more suscseptible to degradation of mechanical properties and environmental resistance with temperature than \textalpha{} \cite{LandW,Rosa1970}.

Alloys known to experience a dwell fatigue life debit include Ti--64, Ti--6246 and Ti--834 \cite{Zheng2018}. The Al$_{\mathrm{eq}}$ values for these alloys are 7.5, 8.9 and 9.3 wt.\%, respectively. Additionally, ageing treatments in the \textalpha{}+\textalpha{}$_2$ field are often used to strengthen the \textalpha{} phase, despite the potential impact on dwell fatigue resistance.

A commonly referenced guideline in titanium alloy design and industry has been that an alloy will be stable and ``unlikely'' to form \textalpha{}$_2$ if Al$_{\mathrm{eq}} \le$ \SI{9}{\wtpercent} \cite{Rosenberg1968,Kosaka2007}. However, this study demonstrates that this is comfortably within the \textalpha{}$_2$-forming regime for the binary Ti--Al and Ti--Sn systems. The mixed ordered phase Ti$_3$(Al,Sn) can also form in alloys containing both solutes \cite{Radecka2016b}. It is further noted that, in an \textalpha{}+\textbeta{} alloy, the \textalpha{} phase will be Al-- or Sn--enriched relative to the alloy composition.

Typical rotor alloys contain up to 0.2 wt\%~O, an Al$_{\mathrm{eq}}$ increment of 2. Here, the solubility limit of 6.2~wt.\% in an alloy containing 0.08~wt\% O implies an Al$_{\mathrm{eq}}$ limit of 7.0 wt\%. This implies that a Ti--$x$Al--4V alloy containing 0.2 wt\% O should contain no more than 5~wt.\% Al. If, in an \textalpha{}+\textbeta{} alloy such as Ti--6Al--4V, a 10\% volume fraction of Al-depleted $\beta$ phase is assumed, then this implies a lower Al$_{\mathrm{eq}}$ limit of 6.5~wt\%, or an Al limit of 4.5~wt.\% once O is accounted for. Thus, even alloys such as Ti--5Al--7.5V (Ti--575) may not be entirely safe from $\alpha_2$ precipitation.

From an alloy design perspective, we therefore suggest that current structural \textalpha{}+\textbeta{} alloys for aerospace applications (especially critical rotating parts subject to dwell fatigue loading) contain too high an aluminium equivalent to avoid deleterious crystallographic ordering.

In summary, SAXS was used to examine the solubility limits for Al and Sn in $\alpha$-Ti against Ti$_3$(Al,Sn) precipitation. The \textalpha{}/\textalpha{}+\textalpha{}$_2$ boundary was found to lie at \SI{6.2}{\wtpercent} (\SI{10.6}{\atpercent}) for the Ti--Al system and \SI{16.9}{\wtpercent} (\SI{7.6}{\atpercent}) for the Ti--Sn system, for alloys containing around 500 ppmw O, in agreement with literature \cite{Murray1990,Namboodhiri1983}. Tensile testing combined with a Hall-Petch analysis showed that, due to the density penalty associated with Sn additions, Al is a slightly more efficient solution strengthener than Sn, even at the solubility limit. Counter to the compositional design of several widely-used \textalpha{}+\textbeta{} and near-\textalpha{} alloys in use today, it is suggested that a `safer' Al$_{\mathrm{eq}}$ limit may be as low as \SI{6.5}{\wtpercent}, or an alloy Al limit of 4.5 wt.\%, compared to the limit of Al$_{\mathrm{eq}}=9$ wt.\% used traditionally.

\section*{Acknowledgements}
\noindent FFW was funded by Rolls-Royce plc and by the EPSRC Centre for Doctoral Training in the Advanced Characterisation of Materials (EP/L015277/1). DD was funded by a Royal Society Indstrial Fellowship and EPSRC (EP/K034332/1). We thank Dr T.W.J. Kwok and Dr I. Bantounas at Imperial College London for assistance with alloy processing, and TIMET for support with composition measurements. This research used resources of the Advanced Photon Source, a U.S. Department of Energy (DOE) Office of Science User Facility operated for the DOE Office of Science by Argonne National Laboratory under Contract No. DE-AC02-06CH11357.

\bibliographystyle{model1-num-names}
\bibliography{sample.bib}

\end{document}